\documentclass[usegraphicx,usenatbib,useapjfonts,apj]{emulateapj}
\usepackage{amssymb}

\begin{document}

\newcommand{\be}{\begin{equation}}
\newcommand{\ee}{\end{equation}}
\newcommand{\M}{\tilde{M}}
\newcommand{\C}{\mathcal{C}}
\newcommand{\rhocr}{\rho_{\mathrm{cr}}}
\newcommand{\Sigcr}{\Sigma_{\mathrm{cr}}}
\newcommand{\Mvir}{M_\mathrm{vir}}
\newcommand{\rvir}{r_\mathrm{vir}}
\newcommand{\dcr}{\delta_\mathrm{cr}}
\newcommand{\crit}{\mathrm{cr}}
\renewcommand{\sun}{\odot}
\newcommand{\Msun}{\rm M_{\odot}}
\newcommand{\n}{{\bf {\hat {n}}}}

\def\plotone#1{\centering \leavevmode
   \epsfxsize=\columnwidth \epsfbox{#1}}
\def\plotancho#1{\includegraphics[width=18cm]{#1}}
  
\title{The thermal Sunyaev-Zel'dovich Signature of Baryons in the Local Universe}

\author{Carlos Hern\'andez-Monteagudo,\altaffilmark{1} Hy Trac,\altaffilmark{2} Raul Jimenez,\altaffilmark{1} Licia Verde\altaffilmark{1}}

\altaffiltext{1}{Department of Physics and Astronomy, University of Pennsylvania, Philadelphia, PA 19104, USA; carloshm@astro.upenn.edu; lverde, raulj@physics.upenn.edu}
\altaffiltext{2}{Department of Astrophysical Sciences, Princeton, NJ 08544, USA; htrac@astro.princeton.edu}

\begin{abstract}
  
  Using cosmological hydrodynamical simulations, we investigate the
  prospects of the thermal Sunyaev-Zel'dovich (tSZ) effect to detect
  the missing baryons in the local universe. We find that at least
  80\% of the tSZ luminosity is generated in collapsed structures, and
  that $\sim$ 70\% of the remaining diffuse tSZ luminosity (i.e.,
  $\sim 15$\% of the total) comes from overdense regions with
  $\delta_{gas}>$10, such as filaments and superclusters. The gas
  present in slightly overdense and underdense regions with
  $\delta_{gas} < 10$, despite making up 50\% of the total baryon
  budget, leaves very little tSZ signature: it gives rise to only
  $\sim$ 5\% of the total tSZ luminosity. Thus, future Cosmic
  Microwave Background (CMB) observations will be sensitive to, at
  best, one half of the missing baryons, improving the current
  observational status, but still leaving one half unobserved.  Since
  most of the tSZ is generated in haloes, we find a tight correlation
  between gas pressure and galaxy number density. This allows us to
  predict the CMB Comptonization from existing galaxy surveys and to
  forecast the tSZ effect from the local structures probed by the Two
  Micron All Sky Survey (2MASS) galaxy catalog.

\end{abstract}


\section{Introduction}

Recent cosmological observations like galaxy surveys (e.g.,2dFGRS,
\citet{twodf}, SDSS, \citet{sdss}), Supernovae (e.g., SNLS,
\citet{SNLS}) or Cosmic Microwave Background (CMB) (e.g., WMAP,
\citet{wmap1,wmap2}) give strong support to the LCDM model of the
universe, in which only $\sim$ 18\% of matter is in the form of
baryons.  Baryons are mostly probed by CMB, Lyman--$\alpha$ forest
(e.g., \citet{croft,McDonald}), and X-rays observations. The first two
probes show that the Universe, at much younger epochs, contained an
amount of baryons that exceeds by a factor of $\sim$ 9 the baryons
seen in today's universe, \citep{fukpeebles}. This is the missing
baryons problem.
\begin{figure*}
\centering
\plotancho{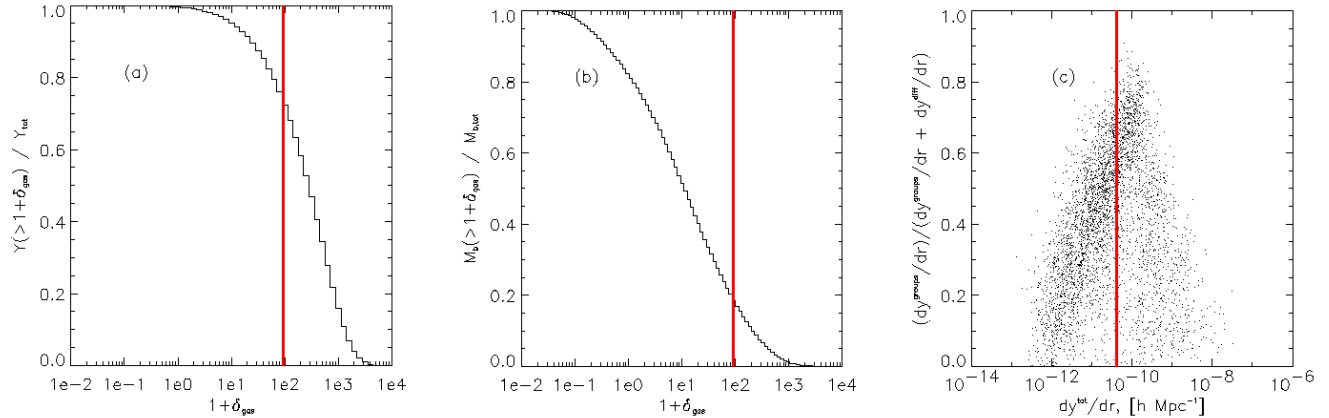}
\caption[fig:fig1]{
  {\it (a)} Relative contribution to the total tSZ luminosity of
  regions with gas density contrast bigger than $\delta_{gas}$. The
  vertical solid line indicates the average gas overdensity in haloes.
  {\it (b)} Relative contribution to the total baryon mass of regions
  with gas density contrast bigger than $\delta_{gas}$. The vertical
  solid line indicates the average gas overdensity in haloes. Note the
  asymmetry with respect to the previous (left hand side) panel.  {\it
    (c)} Ratio of gas pressure produced in galaxy groups (small
  haloes) over pressure produced by diffuse gas {\em and} galaxy
  groups, versus total pressure (including galaxy clusters), averaged
  on scales of 12 h$^{-1}$Mpc. The solid vertical line corresponds to
  regions with average matter density. In overdense regions, the
  pressure of haloes always dominates the diffuse pressure. }
\label{fig:fig1}
\end{figure*}
Indeed, according to \citet{cenost1,cenost2}, about half of the
baryons should be in the form of a Warm Hot Intergalactic Medium
(WHIM), at relatively high temperature ($\sim 10^5-10^7$ K) but low
densities. Only a small fraction of the total gas is hot enough
($^>_{\sim} 10^8$K) to enable X-ray detection via {\it
  bremmstrahlung}: this mechanism led in the nineties to the
identification of massive galaxy clusters from ROSAT X-ray
observations (e.g., \citet{rosat}), and more recently has motivated
the search (and tentative detections) of diffuse X-ray emission from
filaments and superclusters, \citep{xr1,xr2}. Although further pieces
of evidence for local gas have been found in quasar spectra, in the
form of Oxygen absorption features at $z\simeq 0$
\citep{qspec1,qspec2}, it is clear that we do not have a complete map
of the baryon distribution in the local Universe.

With the advent of high resolution and high sensitivity CMB
experiments it has been proposed to trace the local gas using the
thermal Sunyaev-Zel'dovich effect (tSZ, \citet{tSZ}). The tSZ effect
describes the Compton scattering of CMB photons off hot electrons, in
which a net transfer of energy from electrons to radiation distorts
the Black Body spectrum of the CMB. The relative 
intensity fluctuations introduced in this scattering are redshift
independent, and their amplitude is proportional to the integrated
electron pressure along the line of sight.  Observational efforts by
\citet{tenef1} and \citet{mito} have detected strong decrements in
Corona Borealis, and interpreted them as Compton scattering being
generated by extended gas in a filament aligned with the line of
sight. Other studies \citep{hansen05, dolag05, atrio06} addressed this
problem from a theoretical point of view, trying to characterize the
tSZ signal generated by non-collapsed baryons.

Here we use a hydrodynamical numerical simulation to explore the
prospects of future tSZ observations to search for the missing
baryons.  We find that, as the tSZ effect is sensitive to gas
pressure, most of the tSZ luminosity (around 80\%) is generated in
collapsed structures, which host only $\sim$30\% of the total baryon
mass.  Most baryons are in environments with low overdensities, giving
rise to very little tSZ signal: more than 60\% of baryons are in
regions with $\delta_{gas} < 20$, which contribute only $\sim$10\% of
the total tSZ luminosity. More than two thirds of the diffuse tSZ
luminosity (not generated in collapsed structures) is coming from
overdense environments with $\delta_{gas} > 10$ such as filaments or
superclusters. The significant amount of baryons present in underdense
regions ($\sim$ 18\% where $\delta_{gas} <0$) is practically invisible
to the tSZ effect ($ \sim 0.5$ \% contribution to tSZ luminosity).
While the tSZ will be, at best, sensitive to $ \delta_{\rm gas} > 10$
(i.e. superclusters and filaments), which corresponds to $\sim 50\%$
of all baryons, X-rays are only sensitive to the core centers of
galaxy clusters in our neighborhood which corresponds to less than 5\%
of all baryons, \citep{fukpeebles}.

Because in overdense regions most of the tSZ is generated in haloes,
we find a tight correlation between the gas pressure giving rise to
the tSZ and the number galaxy density field, $p_e \propto
n_{gal}^{2.2}$. This allows us to build tSZ templates and to make
predictions for the tSZ signal from existing galaxy surveys.
\section{Numerical Simulations}
We ran a cosmological simulation of a LCDM cosmology ($ \Omega_m=0.3,
\Omega_\Lambda=0.7, \Omega_b=0.045, h=0.7, \sigma_8=0.9, n=1$) with a
box of 200 Mpc/h side, $1024^3$ hydro grid cells, and $512^3$ dark
matter particles using the TVD+PM code of \citet{hy04}.  The comoving
grid spacing is 195 kpc/h and the dark matter particle mass resolution
is $5\times10^9 M_\odot/h$.  The Eulerian hydro algorithm computes the
spatial and temporal changes in the conserved mass, momentum, and
total energy.  The thermal energy is calculated by subtracting the
kinetic energy from the total energy, while the pressure and
temperature are related to the thermal energy by the standard equation
of state. The simulation is adiabatic. We recall the definition of the
the Comptonization parameter ($y\equiv \int dr \sigma_T\;n_e(k_B
T_e)/(m_e c^2)$, with $\sigma_T$ the Thomson cross-section, $k_B$ the
Boltzmann constant and $m_e, n_e, T_e$ the electron mass, density and
temperature, respectively), in order to introduce the derivative
$dy/dr$ as a proxy for electron pressure, ($dy/dr \propto p_e \propto
n_e T_e$).  Likewise, we define $Y \equiv \int d^3r\; dy/dr$, as a
proxy for tSZ luminosity, since both quantities are also proportional.
Bound dark matter particles classified as halos are identified using
an ellipsoidal overdensity threshold of 200 times the average density.
This defines the halo boundaries which are used to distinguish between
tSZ generated in collapsed structures and in the diffuse phase.
\begin{figure*}
\centering
\plotancho{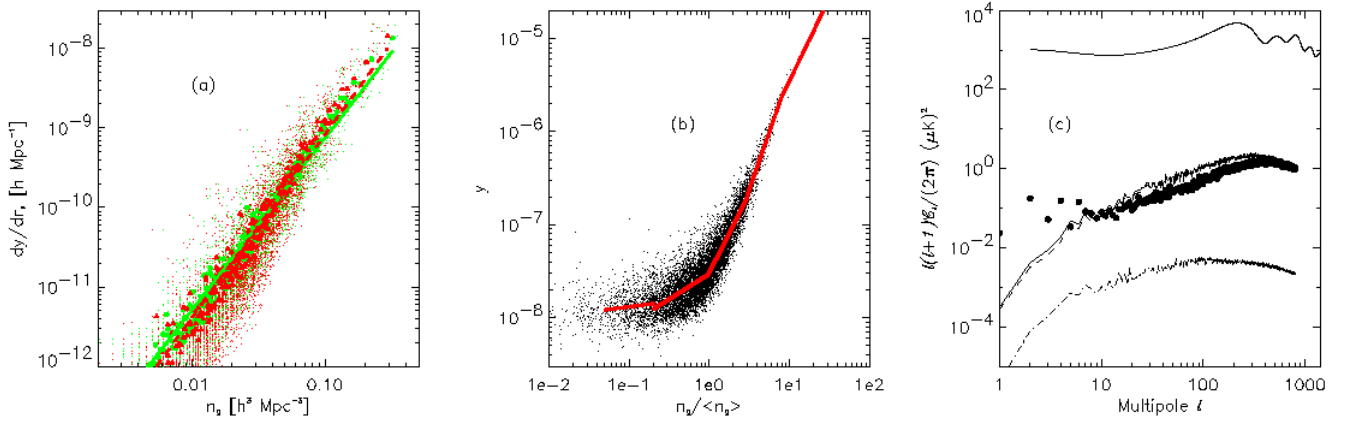}
\caption[fig:fig2]{
  {\it (a)} Pressure -- galaxy number density correlation, according
  to SSHJ (filled circles and solid line) and K04 (filled triangles,
  dashed line) prescription to populate haloes with galaxies. {\it
    (b)} Correlation between Comptonization parameter $y$ and
  projected galaxy overdensity (K04 prescription) obtained after
  projecting a simulated sphere of of 300 h$^{-1}$ Mpc radius. {\it
    (c)} Rayleigh-Jeans (RJ) expectation for our prediction of the
  local tSZ power spectrum computed from 2MASS galaxy catalog and the
  $y$--galaxy correlation (filled circles). The tSZ power spectrum
  (RJ) from our simulated volume is given by the solid line, and
  decomposed into the halo (dashed line) and diffuse (dot-dashed line)
  contributions.  }
\label{fig:fig2}
\end{figure*}
\section{Where is the tSZ produced?}
We distinguish tSZ luminosity being generated in cells belonging to
{\it clusters of galaxies} ($M_{cl} \geq 5\times 10^{13}\;h^{-1}
M_{\odot}$), to {\it small} haloes (also referred to as {\it galaxy
  groups}, and defined as all haloes resolved in the simulation with
masses below $5\times 10^{13}\;h^{-1} M_{\odot}$), and to a {\it
  diffuse} gaseous phase (defined as all gas cells not belonging to
any halo). Given the dark mass particle resolution in our simulation,
we choose our mass threshold for the halo definition to be $\sim
10^{12}\;h^{-1} M_{\odot}$: all haloes below this limit are
regarded as diffuse gas, and hence our estimates for the diffuse tSZ
contribution should be regarded as optimistic.

Figure (\ref{fig:fig1}a) shows the cumulative distribution of the
pressure in the box versus gas density contrast.  Collapsed structures
show a halo-mass-weighted average gas overdensity of $\sim 93$,
(marked by the vertical solid line).  When integrating -i.e., volume
weighting- the tSZ luminosity in the three cell subsets defined above
(galaxy clusters, small haloes and diffuse gas), we find that $\sim$
70\% of total tSZ luminosity is generated in galaxy clusters, and, out
of the $\sim $ 30\% remaining, around one third of it (i.e., $\sim$
10\% of the total) is generated in small haloes.  This leaves the
diffuse phase a $\sim$ 20\% contribution, even though it hosts
$\sim$70\% of the total baryonic mass. Further, one fourth of this
diffuse gas is located in underdense regions, whose tSZ luminosity is
negligible ($<$ 1\% ). This asymmetry of the pressure and baryon mass
distributions versus gas overdensity is explicitely shown in Figures
(\ref{fig:fig1}a,b).

This suggests that the detection, via the tSZ effect, of the diffuse
gas will be hampered by the presence of small (and plausibly
unresolved) haloes: this is shown in Figure (\ref{fig:fig1}c), where
the ratio of the gas pressure generated in small haloes over the sum
of the contributions from the diffuse gas and small haloes is plotted
versus the total gas pressure, (in all cases the pressure is computed
in cells of 12 $h^{-1}$Mpc side).The vertical line marks the
total pressure corresponding to the average galaxy number density.  We
see that, in slightly overdense regions hosting large pressure 
(right of the vertical line), the tSZ luminosity outside
galaxy clusters is preferentially generated in smaller haloes rather
than in diffuse gas, i.e., when searching for the tSZ signature of
diffuse gas in overdense regions one must carefully subtract the
contribution from small haloes.  Projection effects in future CMB maps
will further enhance this contamination. Only when considering the
very overdense regions of the universe (far right end of this plot),
the relative weight of the diffuse phase takes over, since in such
environments there are no small haloes (all haloes are clusters). Note
however that the mass threshold for our cluster definition ($5\times
10^{13}\;h^{-1}M_{\odot}$) corresponds to abundant and
not-so-tSZ-bright objects, and that these sources may still be
responsible for a significant amount of confusion noise.
\begin{figure*}
\plotancho{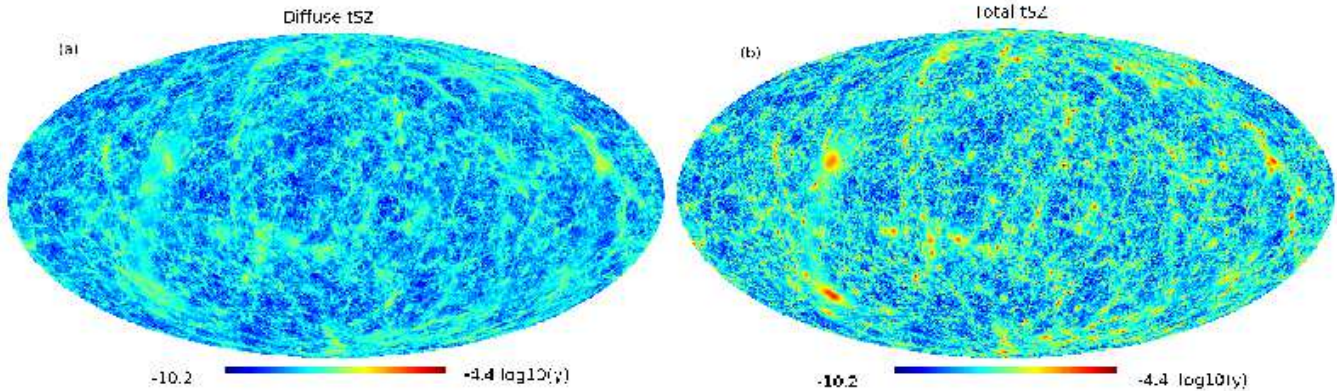}
\caption[fig:fig3]{
  {\it (a)} All-sky diffuse tSZ contribution in our simulated volume, as oppossed to the total tSZ contribution, (panel {\it (b)}).
}
\label{fig:fig3}
\end{figure*}
\section{The pressure - halo correlation}
Since more than 80\% of the tSZ luminosity is generated in collapsed
structures, there must be a correlation between the gas pressure and
the number density of haloes, or the gas pressure and the galaxy
number density.

To investigate this correlation, we populate the haloes in the
simulation using two different algorithms: the first one
(\citet{scoccimarro}, hereafter SSHJ) characterizes the halo galaxy
population by a binomial distribution, whereas the second
(\citet{kravtsov}, hereafter K04) uses a Poissonian distribution. As
in \citet{smith05}, both algorithms are normalized to give the same
average galaxy number density, ($\langle n_g \rangle \simeq 3 \times
10^{-2}\; h^3$Mpc$^{-3}$), and this requires slightly different
choices for the halo minimum mass hosting a galaxy.  As shown in
Figure (\ref{fig:fig2}a), the correlation found between $dy/dr$ and
galaxy number density (computed in cells of 12$h^{-1}$Mpc side) is
very similar in both cases, and well fitted by a power law of index
$n\approx 2.1-2.3$ for the SSHJ and K04 methods (filled circles and
filled triangles, solid and dashed lines, respectively).
The slope obtained in both cases is close to the prediction that a
polytropic gas model ($p \propto \rho^{\gamma}$) provides for a
self-gravitating system, ($\gamma = 2$).

As current all sky maps of the local galaxy distribution do not have
distance information, we want to investigate the correlation between
projected galaxy density and $y$.  We extend our simulated volume to a
box of 600 h$^{-1}$Mpc side, exploiting the fact that the simulation
box has periodic boundary conditions.  We then place an observer in
the center and project on a HEALPix \citep{healpix} 
 map of resolution parameter
N$_{side}$ = 512 a sphere of 300 h$^{-1}$Mpc radius. We produce maps
of $y$ and galaxies, smooth them with a Gaussian beam of FWHM = 12.6
arcmins, and sort the pixels in increasing projected number of
galaxies. After binning them in groups of 32 pixels (for display
purposes), we obtain a correlation between $y$ and the projected
galaxy overdensity ($n_g (\n ) / \langle n_g \rangle$, $\n$ denoting a
direction on the sky), shown in Figure (\ref{fig:fig2}b).

We find that for slight galaxy overdensities (few times the background
density), there is a tight correlation between galaxy number density
and the Comptonization parameter $y$. This suggests using existing
galaxy catalogs to predict the degree of Comptonization of the CMB
sky. The volume used in our projection is close to that sampled by the
Two Micron All Sky Survey (hereafter 2MASS, \citet{2mass}). We thus
use a map of projected galaxies from this survey (constructed as in
\citet{chmrgsfab}) to make a prediction of the $y$ sky. When comparing
the galaxy maps from 2MASS and our simulation we find that they both
show similar galaxy densities and clustering properties. We
approximate the correlation given in Figure (\ref{fig:fig2}b) by five
power laws (given by the solid lines), and invert the 2MASS-based
galaxy catalog into a $y$ map.

The overall $y$ normalization is set by imposing that the 256 highest
density pixels have a tSZ decrement of -73 $\pm$ 17 $\mu$K, as found
at 94 GHz by \citet{chmrgsfab} (all maps have been convolved by the
same Gaussian beam). The power spectrum of the resulting $y$ map (in
units of [$\mu$K]$^2$ at Rayleigh-Jeans frequencies) is shown in
Figure (\ref{fig:fig2}c) by filled circles: its amplitude and shape is
very close, at high multipoles, to the power spectrum of the $y$ map
produced from our simulation (solid line). They only differ clearly at
large scales ($\ell < 10$), for which the $y$ power spectrum estimated
from 2MASS is higher. This prediction for the local tSZ
power spectrum is remarkably close both in shape and amplitude to
those obtained by \citet{hansen05, dolag05} via a independent
approach. Our computation  should  be a valid prediction
for the tSZ power at low $\ell$s, but at high $\ell$s the neglected high-redshift cluster
population will dominate.
\section{Prospects for diffuse tSZ detection}
Since we have separated in our simulation the gas belonging to
collapsed haloes from the gas in a diffuse phase, it is possible to
compute $y$ maps and power spectra of the halo and the diffuse
components separately.  The dot-dashed line at the bottom of Figure
(\ref{fig:fig2}c) provides the power spectrum of the diffuse gas in
our simulation, whereas the dashed line corresponds to the tSZ power
spectrum generated by gas located in haloes (note its proximity to the
total contribution [solid line]). A map of diffuse tSZ is given in
Figure (\ref{fig:fig3}a), and comparing to the total tSZ contribution
(Figure (\ref{fig:fig3}b)). We conclude that most of the diffuse tSZ
signal visible on the CMB sky (and more than 90\% of the total tSZ
luminosity) is generated in overdense regions traced by the clusters
and superclusters, (i.e., the diffuse warm gaseous component present
in voids leaves negligible tSZ signature). The amplitude of the
diffuse phase is at best ten times (a hundred times in $C_{\ell}$'s)
smaller than the tSZ signal generated in haloes, {\em even at large
angular scales}.  In particular since galaxies trace superclusters and
filaments (corresponding to $\delta_{gas}>10$), the diffuse gas in
these regions could be targeted by tSZ observations with
specifications similar to those of e.g. ACT ({\tt
http://www.hep.upenn.edu/act/ }). These regions contain 50\% of the
baryons: thus tSZ observations may decrease the ratio of unseen
baryons in the local universe from $\sim 9$ to $\sim 2$. In practice,
the limiting factor in detecting this signal will be confusion noise
from unresolved haloes.  Note that the ratio of tSZ luminosities of
the collapsed and diffuse baryon components does not equal the ratio
of the corresponding tSZ-induced angular power spectra, since the
latter quantities depend on the contrast of tSZ sources against the
CMB background. This makes the search for the (diffuse) missing
baryons even more challenging. The detection of part of the diffuse
tSZ may be feasible in nearby superclusters after masking out all
compact sources and well characterized nearby galaxy clusters.
However, we must conclude that the tSZ, despite of providing a new
tool to unveil the presence of unseen warm and hot gas, will still
miss a significant fraction of the baryons in the local Universe.

\acknowledgments We thank R.E.Smith and R.Sheth for useful
discussions. We acknowledge the use of HEALPix \citep{healpix}
package, and the Two Micron All Sky Survey.  CHM is supported by NASA
grants ADP03-0092 and ADP04-0093 and NSF grant PIRE-0507768. RJ is
partially supported by NSF grant PIRE-0507768 and by NASA grant
NNG05GG01G. LV is supported by NASA grants ADP03-0092 and ADP04-0093.




\end{document}